\documentstyle[prl,aps]{revtex}
\begin{document}

\input {psfig.tex}

\title{
\bf \Large X-ray emission during the muonic cascade in hydrogen}

\author{B. Lauss,\cite{Email} P.Ackerbauer, 
W.H.Breunlich, B.Gartner, M.Jeitler,
P.Kammel,\cite{adr} J.Marton, W.Prymas, and J.Zmeskal}
\address{Institute for Medium Energy Physics,
Austrian Academy of Sciences,
Boltzmanngasse 3, A-1090 Wien, Austria}

\author{D.Chatellard, J.-P.Egger, and E.Jeannet}
\address{Institut de Physique de l'Universit\'e, 
CH-2000 Neuch\^atel, Switzerland}

\author{H.Daniel, F.J.Hartmann, and A.Kosak}
\address{Physics Department, TU M\"unchen, D-85747 Garching, Germany}

\author{C.Petitjean}
\address{Paul Scherrer Institut, CH-5232 Villigen, Switzerland}

\date{\today}

\maketitle

\begin{abstract}

We report our investigations
of X rays emitted during the muonic
cascade in hydrogen employing charge coupled devices
as X-ray detectors. The density dependence
of the relative X-ray yields for the muonic hydrogen lines (
$\rm K_{\alpha}$, $\rm K_{\beta}$ and
$\rm K_{\gamma}$) has been measured
at densities between 0.00115 and 0.97 of liquid hydrogen density.
In this density region collisional processes dominate 
the cascade down 
to low energy levels. 
A comparison with recent calculations is given in order to
demonstrate the influence of Coulomb deexcitation. 

\end{abstract}

\pacs{PACS numbers: 36.10.Dr, 32.30.Rj, 32.70.Fw }

Exotic atoms are atoms in which an electron is replaced by a
heavier negatively charged particle, e.g. $\rm \mu^{-}, 
\pi^{-}, K^{-},
\overline{p}$.
The simplest of these atoms is muonic hydrogen $\rm \mu p$,
which denotes a bound state of one proton and one muon.

After a free muon is injected into a hydrogen target 
it is slowed
down and an excited muonic hydrogen atom is formed via 
Coulomb capture
\cite{Leon62,Coh83}, leaving the muon most likely in 
an initial state of
$\rm 11 \le n \le 15$ \cite{Korenman}.
This is the starting point for a complicated interplay 
of competitive
collisional and radiative deexcitation processes,
the so-called atomic cascade, which ends with the
muonic atom being in the 1s ground state.
The possibility of a metastable 2p state in muonic hydrogen 
is currently under investigation \cite{Taqqu}.

Muons, in contrast to the other possible particles, 
are not affected by
strong interaction. So they may serve as the best probe for the 
investigation of these deexcitation processes.

At the investigated hydrogen densities, collisional deexcitation 
dominates
from the beginning of the cascade to low n states, 
depending on the target density
and the kinetic energy of the $\rm \mu p$ atom
\cite{Leon62,Leon71,Marku81,Marku94,Asch96}.
The known collisional processes are chemical dissociation,
elastic collision, external Auger electron emission,
Coulomb deexcitation and Stark mixing.
The importance of the last two effects is
due to the very small size of the neutral $\rm \mu p$ atom,
which can penetrate very deeply into the electron cloud of
a neighboring $\rm H_{2} $ molecule, where it can ``feel'' the
electromagnetic field of the protons.

In the investigated hydrogen gases,
the transitions to the ground state are
always accompanied by the emission of muonic X rays;
at liquid hydrogen density (LHD)$\rm ^{\ddag}$,
the X-ray emission probability is still 0.95 \cite{Marku94}.

All these processes together make up the ``standard cascade model''
\cite{Leon62,Leon71,Marku81,Leon80,Kottmann}
which, despite its success, has one severe limitation: this is
the assumption that
$\rm \epsilon_{\mu p}$, the kinetic energy 
of the $\rm \mu p$ atom, is constant throughout the cascade.

A recent calculation \cite{Marku94,Asch96} tries
to take into account the time evolution
of $\rm \epsilon_{\mu p}$ for energy levels n $\rm \le$ 6.
Due to Coulomb deexcitation,
complex energy distributions at various levels are expected.
Coulomb deexcitation is the only known mechanism which
can accelerate $\rm \mu p$ atoms up to epithermal energies
of $\rm \sim 170$ eV \cite{Marku94}.
In spite of its importance,
Coulomb deexcitation is regarded as
the least known process of the muonic
cascade (neglecting chemical deexcitation, which is important
only at very high energy levels) \cite{Asch96}.
Existing calculations of the 
Coulomb deexcitation cross sections
differ among themselves by more than one order of magnitude
\cite{Bra78,Mensch,Kravtsov,Czap95,Pono96}.

Apart from the basic interest in exotic atoms,
the understanding of the deexcitation of muonic atoms 
is of great importance for various other phenomena, such as 
muon catalyzed fusion \cite{Annu,Zme90,Kammel85},
muon transfer \cite{Lau97,Lau95,Gar95},
diffusion of muonic atoms \cite{Abb97}, and nuclear muon capture 
\cite{Bre80,Bre81}. The muonic case 
can also be used as a test for the cascade in hadronic
atoms \cite{Hayano} (e.g. $\rm \pi^{-}$p)
where velocity effects \cite{Bader97,Asch95} 
severely affect the evaluation of strong interaction parameters
\cite{Chat95,Sigg95}.

For this measurement,
the relative X-ray yields of the emitted muonic hydrogen 
$\rm K^{\mu p}$ lines 
-- the corresponding X-ray energies are 
$\rm K_{\alpha}$ = 1.90 keV,
$\rm K_{\beta}$ = 2.25 keV, $\rm K_{\gamma}$ = 2.37 keV
and $\rm K_{\infty}$ = 2.53 keV --
and their intensity ratios 
serve as a tool to investigate the muonic cascade.

This is the first systematic investigation of 
muonic X rays in this density region, and the first one
which gives intensities for the three lowest muonic K lines.
Earlier experiments \cite{Placci,Budick,Ander77,Egan,Ander85}
were either rather inaccurate or
performed at very low gas densities ($\rm < 10^{-3}$ x LHD) only.

Our measurements were carried out with the high intensity muon 
beam of the 
$\rm \mu$E4 area at PSI (Paul Scherrer Institut, 
Villigen, Switzerland).
Figure 1 shows the setup for the measurements at liquid hydrogen 
density. 
Several different silver-coated steel or aluminum target cells 
were optimized
for the measurements in liquid and gaseous hydrogen, respectively
\cite{Lau97b}. 

To minimize X-ray absorption, Kapton windows with a thickness of
12.5 $\rm \mu m$ were
used for the measurements at LHD.
The gas target equipped with 25 $\rm \mu m$ thick windows 
had to withstand pressures of up to 6 bar at
temperatures around 30 K. For safety reasons,
the target vacuum vessel was separated from the CCD's vacuum with an 
additional 12.5 $\rm \mu m$ Kapton window.
The target cells were partly covered with superinsulation
to reduce radiation heating.

\begin{figure}[hbt]
\centerline{\psfig{file=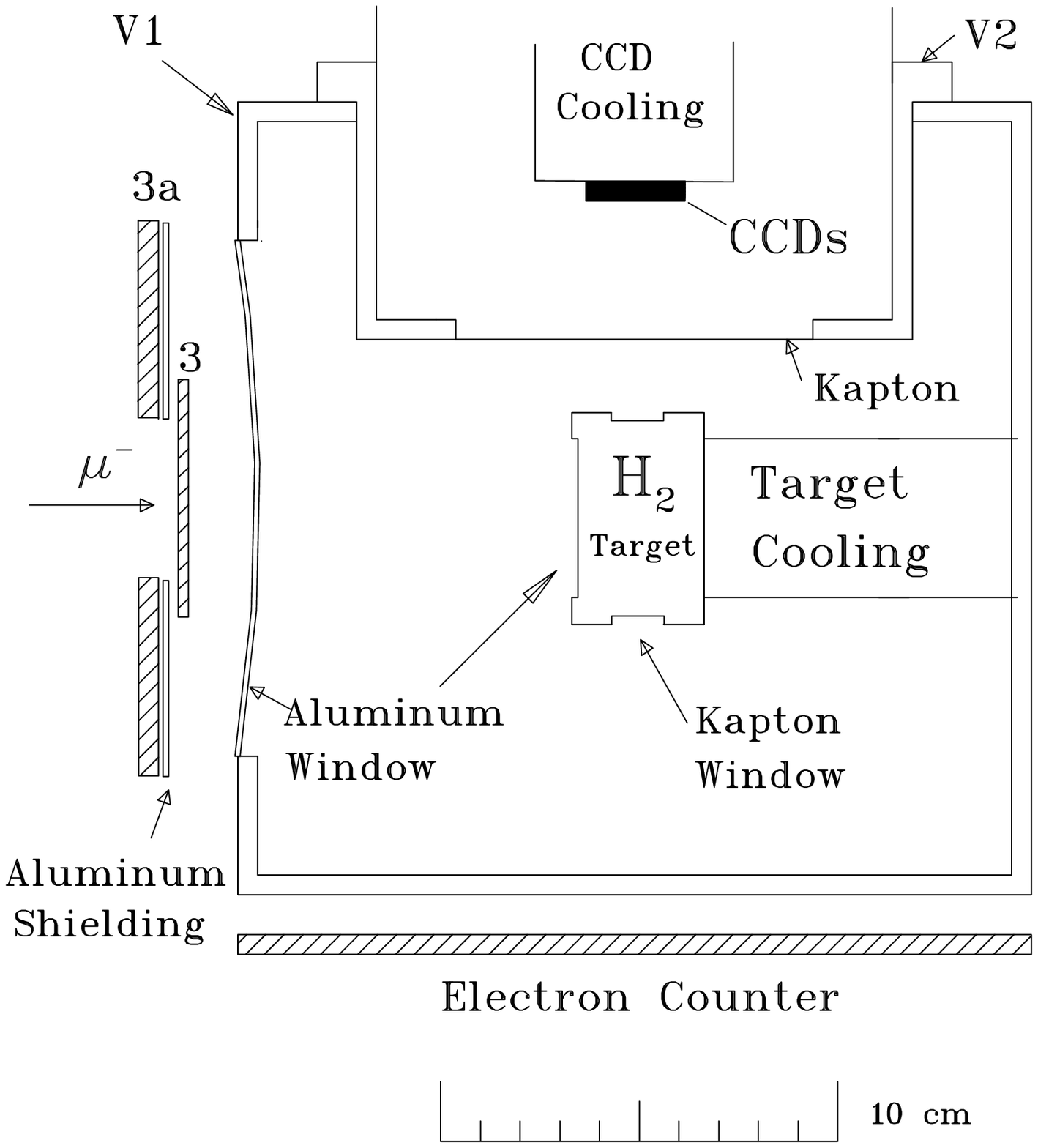,height=7.5cm}}
\caption{ 
Top view of the setup of this experiment. The displayed 
target cell was
used for the measurements in liquid hydrogen.
Two separated vacuum vessels were used; one 
for the target (V1), and one for the CCD-detector (V2). To ensure 
an optimal
stopping efficiency the scintillation counters 
3 and 3a were used to register incoming muons.
The electron counter served to detect muon decay electrons.
Larger target cells,
adapted to the expected extent of the muon stopping distributions,
were used for the measurements in gaseous 
hydrogen \protect \cite{Lau97b}.
}
\label{FIG.1}
\end{figure}

To avoid high Z impurities in the target, which would change the 
K line intensities through excited-state muon transfer 
\cite{Gershtein,Lau97},
the hydrogen was cleaned
during the filling procedure by using a liquid-nitrogen trap 
together with
a palladium filter.
The composition of the hydrogen was checked online 
during the measurements by using a quadrupole mass spectrometer
\cite{Gar93} which could extract samples via a small capillary 
leading directly into the target volume.
The stability of the target pressure and temperature was 
monitored and controlled during all measurements.

Charge coupled devices (CCDs)\cite{Var90a,Var90b} have been 
employed as X-ray detectors.
They consisted of two CCD sensors$\rm ^{\S}$ 
with an active detection area of
$\rm \sim$ 25 x 17 $\rm mm^{2} $($\rm \sim 880000$ pixels) for 
each chip.   
The main chip component was silicon with small absorption 
layers of
$\rm SiO_{2}$ and $\rm Si_{3}N_{4}$ on the surface. 
The depletion thickness
of $\rm \sim$ 30 $\rm \mu $m was well adapted for the observed 
energy region.
To shorten the readout time, each chip was split into 
two electronically
independent detection areas.

\begin{figure}[hbt]
\centerline{\psfig{file=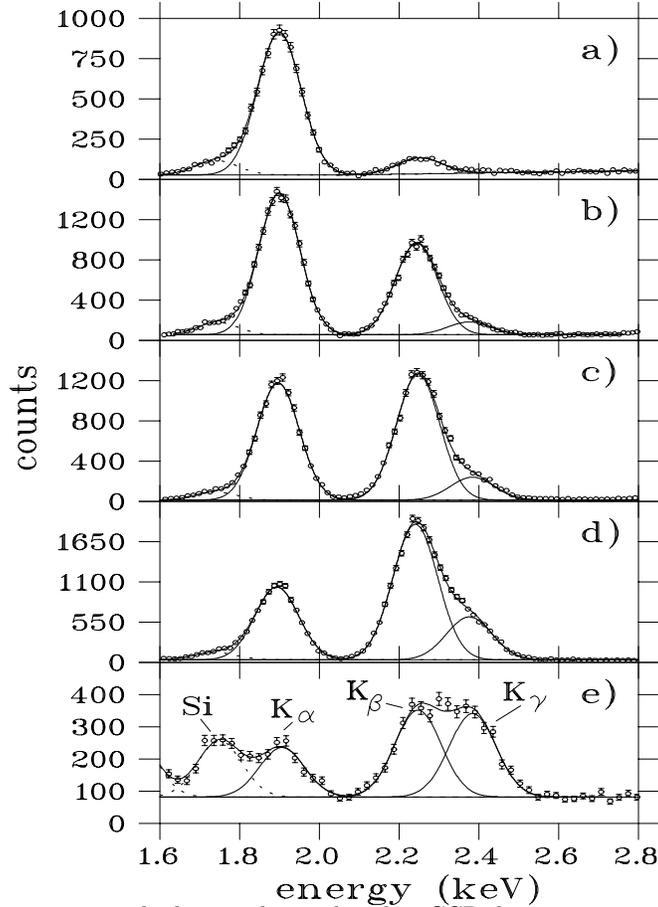,height=12cm}}
\caption{
Muonic X-ray energy spectra in hydrogen observed with a 
CCD-detector
at temperatures of $\rm \sim$ 30 K
and various target densities $\rm \Phi$ (given in units of LHD): 
a) $\rm \Phi  = 0.97$, b) $\rm \Phi  = 0.0795$, 
c) $\rm \Phi  = 0.0391$, d) $\rm \Phi  = 0.0106$, 
e) $\rm \Phi  = 0.00115$.
\protect $\rm K_{\alpha}$, $\rm K_{\beta}$ and 
\protect $\rm K_{\gamma}$ lines
could be separated. The energy resolution of the detector 
$\rm \Delta E^{FWHM}/ E$ at 2 keV was 6 $\rm \%$.
The solid lines indicate Gaussian fits.
The density dependence of the line intensities is clearly visible.
The X-ray peak at 1.74 keV (dashed line) is due to fluorescence 
excitation of the detector's silicon material.
The low counting statistics in measurement e) 
reflects the already very low stopping probability of muons
in such a low-density hydrogen gas. 
}
\label{FIG.2}
\end{figure}

We were able to apply a specific pattern recognition algorithm to 
separate ``true'' X-ray hits from charged particles and cosmic 
background
using a ``single pixel'' selection criterion 
\cite{Egger93a,Egger93b,Sigg94}. 
A ``single pixel'' was considered to be a ``true'' X ray if 
the charge content 
of the surrounding eight neighbor pixels was statistically 
compatible
with the noise peak of the CCDs.

Data runs lasted for three minutes to guarantee that not 
more than 
$\rm \sim$ 15 $\rm \%$ of the CCD's pixel were hit. A 
longer exposure
time would have caused a decrease in the detection efficiency.
A fraction of hit pixels of more than 30 $\rm \%$ actually 
would have
made it impossible to apply our selection criterion for X rays.

Our measurements investigated a density region covering three 
orders of magnitude. The observed raw energy spectra are displayed 
in fig.2. The intensity variation of the muonic hydrogen 
$\rm K_{\alpha}$, $\rm K_{\beta}$ and $\rm K_{\gamma}$ lines 
with target density can be seen directly. 
No K lines higher than $\rm K_{\gamma}$ could be observed.

\begin{figure}[tbh]
\centerline{\psfig{file=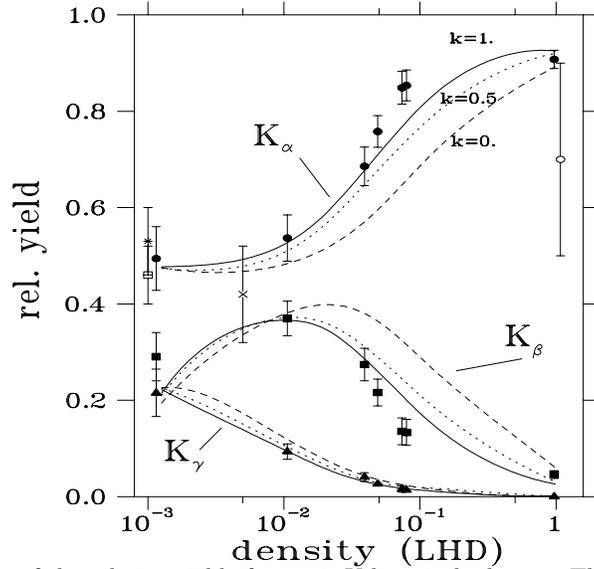,height=7.5cm}}
\caption{
The density dependence of the relative yield of muonic K lines 
in hydrogen.
The measured relative yield is given as the number of counts in one
K line ($\rm K_{i}$, i=$\rm \alpha, \beta, \gamma$) divided by the 
total count rate in all observed K lines
($\rm K_{tot}$=$\rm K_{\alpha}$+$\rm K_{\beta}$+$\rm K_{\gamma}$).
The plotted data points
at LHD include a 5 $\rm \%$ correction for non-radiative 
ground state transitions \protect \cite{Marku94}.
Filled circles, squares and triangles
indicate the relative yield for $\rm K_{\alpha}$,
$\rm K_{\beta}$ and
$\rm K_{\gamma}$ transitions, respectively.
The other experimental points for $\rm K_{\alpha}$ yields
are taken from Refs.
\protect \cite{Placci}(cross), 
\protect \cite{Budick}(open circle),
\protect \cite{Ander77}(star) 
and \protect \cite{Egan} (square).
The displayed calculated results from 
\protect \cite{Asch96} used
a scaling factor k for the Coulomb deexcitation cross section;
k = 0 (dashed line), k = 0.5 (dotted line) and k = 1 (full line).
}
\label{FIG.3}
\end{figure}

An empty target run and a $\rm \mu^{+}$ run proved that there 
were no 
background peaks visible within the relevant energy region.
The X-ray peak at 1.74 keV 
is due to fluorescence excitation of the detector's 
silicon material.
Therefore the background function could be approximated by the 
sum of
a constant and a term depending linearly on energy. 
Gaussians with
energy-dependent width \cite{Lau97b,Egger93a} were used
to fit the peak areas.

The knowledge of the X-ray detection efficiency was 
indispensable for a correct analysis. 
A Monte Carlo program was written \cite{Lau97b} to 
correctly account 
for the various contributions to X-ray absorption. The 
geometry of the different target cells, the absorption in 
the windows and in the CCDs' top layers was simulated. 
The intrinsic detection efficiency of the CCDs 
and the absorption in the Kapton foils
were measured experimentally \cite{Lau97b}.

\normalsize

\begin{center}
\begin{table}
\label{TAB.I}
\caption{Results of the muonic X-ray measurements in hydrogen.}
\begin{tabular}{ccccc} 
{\bf Density [LHD]} & $\rm \bf K_{\alpha}/K_{tot}$ & 
$\rm \bf K_{\beta}/K_{tot}$ & $\rm \bf K_{\gamma}/K_{tot}$ &
 {\bf $\rm \bf K_{\alpha}/K_{\beta}$} \\     \hline
 0.9700 $\rm \pm$ 0.0050    & 0.952 $\rm \pm$ 0.019  &
 0.048 $\rm \pm$ 0.008 & 0.000 $\rm \pm$ 0.010 &
 19.92 $\rm \pm$ 2.45 \\    \hline
 0.0795 $\rm \pm$ 0.0008    & 0.854 $\rm \pm$ 0.032  &
 0.133 $\rm \pm$ 0.027 & 0.013 $\rm \pm$ 0.006 &
  6.41 $\rm \pm$ 1.25 \\ \hline
 0.0738 $\rm \pm$ 0.0008    & 0.849 $\rm \pm$ 0.034  &
 0.135 $\rm \pm$ 0.028 & 0.016 $\rm \pm$ 0.007 &
  6.27 $\rm \pm$ 1.75 \\ \hline
 0.0489 $\rm \pm$ 0.0004    & 0.758 $\rm \pm$ 0.033  &
 0.216 $\rm \pm$ 0.028 & 0.026 $\rm \pm$ 0.007 &
  3.51 $\rm \pm$ 0.52 \\ \hline
 0.0391 $\rm \pm$ 0.0003    & 0.686 $\rm \pm$ 0.040  &
 0.274 $\rm \pm$ 0.034 & 0.040 $\rm \pm$ 0.009 &
  2.50 $\rm \pm$ 0.42 \\ \hline
 0.0106 $\rm \pm$ 0.0001    & 0.539 $\rm \pm$ 0.048  &
 0.369 $\rm \pm$ 0.036 & 0.092 $\rm \pm$ 0.016 &
  1.46 $\rm \pm$ 0.25 \\ \hline
 0.00115 $\rm \pm$ 0.00005  & 0.494 $\rm \pm$ 0.041  &
 0.291 $\rm \pm$ 0.050 & 0.215 $\rm \pm$ 0.049 &
  1.70 $\rm \pm$ 0.38 \\ 
\end{tabular}
\end{table}
\end{center}

Figure 3 displays our results, given in Tab.I 
for the relative X-ray yields of the muonic hydrogen
$\rm 2 \rightarrow 1$, $\rm 3 \rightarrow 1$ and 
$\rm 4 \rightarrow 1$
transitions. The relative yield is given by the 
intensity of one 
specific K line ($\rm K_{i}$, i=$\rm \alpha, \beta, \gamma$) 
divided by the intensity of all observed K lines
($\rm K_{tot}$ = $\rm \sum_{i} K_{i}$). 
The plotted data at LHD include a 5 $\rm \%$ correction
for non-radiative ground state transfer according to
\cite{Marku94}.
This is the first measurement which allows us to test the
existing cascade calculations at high target densities
\cite{Leon71,Marku94,Asch96,Leon80}. 
The calculations taken from \cite{Asch96}
shown in fig.3 and fig.4 were 
done for different contributions of Coulomb deexcitation.
The corresponding cross sections \cite{Bra78} were
scaled with a factor k = 0 (no contribution / dashed line),
k = 0.5
(dotted line) and k = 1 (full line).
The calculation according to the standard cascade 
model \cite{Leon71,Leon80}
coincides with the k = 0 assumption. 
The displayed data -- error bars
include statistical errors and contributions due to efficiency 
corrections -- match quite well the calculated 
density dependence.
Though not very sensitive to k within the experimental errors,
the yield measurements favor values of k $\rm \approx$ 1.

\begin{figure}[htb]
\centerline{\psfig{file=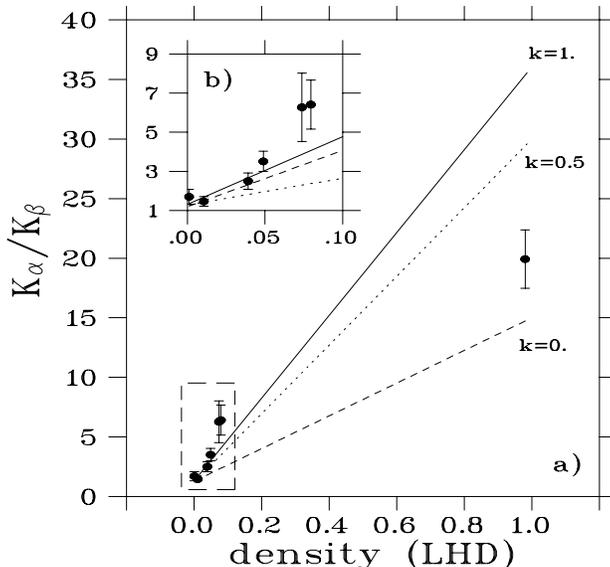,height=7.5cm}}
\caption{
a) The density dependence of the \protect 
$\rm K_{\alpha}/K_{\beta}$ ratio 
in muonic hydrogen. The theoretical curves are calculated as
in fig.3. b) displays the magnified low density region.
}
\label{FIG.4}
\end{figure}

Theory \cite{Asch96} predicts that 
in the observed density region the density dependence of the
$\rm K_{\alpha}/K_{\beta}$ ratio is approximately linear, with
a significant contribution from Coulomb deexcitation.
Figure 4 displays our results (Tab.I) together with 
the calculation 
from \cite{Asch96}. The $\rm K_{\alpha}/K_{\beta}$ value at 
LHD favors a scaling factor k $\rm < $ 0.5.

Although our measurements indicate that Coulomb 
deexcitation plays
a significant role during the muonic cascade
an unambiguous decision on the correct value of
the scaling factor k is not yet possible.

The calculation of the relative line yields
already seems to be quite reliable. However, the 
slight disagreement
at 8 $\rm \%$ of LHD could be an indication for a 
more complex
density dependence of the collisional processes 
in the high density region;
e.g. the existence of possible molecular 
effects \cite{Frolich}.
Using a new calculation of the Coulomb deexcitation
cross sections \cite{Pono96} in the cascade model
could also explain the observed behavior and abolish 
the necessity of the scaling factor introduced 
in \cite{Asch96}.


Financial support by the Austrian Science Foundation, 
the Austrian
Academy of Sciences, the Swiss Academy of Sciences, the Swiss
National Science Foundation and the
German Federal Ministry of Research and Technology 
is gratefully
acknowledged. 
It is a pleasure to thank D.Sigg for his software support and
D.Varidel for his excellent technical assistance. 
Many fruitful discussions with E.C.Aschenauer and V.E.Markushin, 
who provided us with the results of their calculations prior 
to publication, are acknowledged.

\end{document}